\documentclass[aps,pra,twocolumn,amssymb,amsmath,floatfix]{revtex4}
\usepackage{graphicx}
\begin{document}
\title{A Balanced Truncation Primer}
\author{Benjamin Rahn}
\email{brahn@caltech.edu}
\affiliation{Institute for Quantum Information, California Institute of
Technology}
\date{December 11, 2001}

\begin{abstract}
Balanced truncation, a technique from robust control theory, is a
systematic method for producing simple approximate models of complex
linear systems.  This technique may have significant applications in
physics, particularly in the study of large classical and quantum
systems.  These notes summarize the concepts and results necessary to
apply balanced truncation.
\end{abstract}

\maketitle

\def\expval[#1]{\langle #1 \rangle}
\def\tr{\mathrm{tr}}
\def\comment{\bf \footnotesize}
\def\ket[#1]{\vert #1 \rangle}
\def\bra[#1]{\langle #1 \vert}
\def\braket[#1,#2]{\langle #1 \vert #2 \rangle}
\def\ketbra[#1,#2]{\vert #1 \rangle \langle #2 \vert}
\newcommand{\wt}[1]{\widetilde{#1}}

\section{Introduction}
Theoretical physics endeavors to produce models that predict observed
physical phenomena.  Though sometimes the challenge is to develop a
mathematical language describing the system of interest, often
(especially in the study of complex systems) one can write down the
exact dynamics and gain little insight --- the resulting expressions are
too cumbersome, too messy, or too ill-conditioned to be useful.  By
exploiting symmetries and other characteristics of the particular
system, one may find a simpler equivalent description of the dynamics.
This description may be further simplified by using approximation
techniques, e.g., asymptotic limits and small-parameter expansions.
Much of the art of the field lies in finding and choosing ad hoc methods
for deriving these simpler models; however, more systematic methods are
clearly desirable.

Control theorists have developed a variety of \textit{model reduction}
techniques that systematically produce simple models of complex systems.
These notes will describe \textit{balanced truncation} \cite{dullerud,
zhou}, a model reduction technique for linear systems which is readily
available in a variety of formats (e.g.~MATLAB).  Balanced truncation
has recently been applied in physics contexts \cite{qecc_mr,Sznaier},
and the results suggest it will prove a useful tool for treating large
systems in both classical and quantum settings.

Much of this discussion comes directly from Dullerud and Paganini
\cite{dullerud}.  We will present the important concepts and results
necessary to apply balanced truncation, omitting both the proofs and the
algorithms.  We refer the reader to \cite{dullerud} and \cite{zhou} for
a more complete mathematical discussion, and to MATLAB toolboxes and
their documentation \cite{algorithms} for the computational methods.

In section \ref{sec:inout} we will describe the input-output paradigm of
control theory, and introduce state-space models, the class of systems
treatable by balanced truncation.  Given an arbitrary state-space model,
we will characterize the smallest state-space model with identical
input-output characteristics in section \ref{sec:minreal}, and in section
\ref{sec:baltrunc} we will show how balanced truncation is used to find
smaller models with controlled approximation errors.  

\section{Input-Output Maps and State-Space Models\label{sec:inout}}

\begin{figure}
\includegraphics{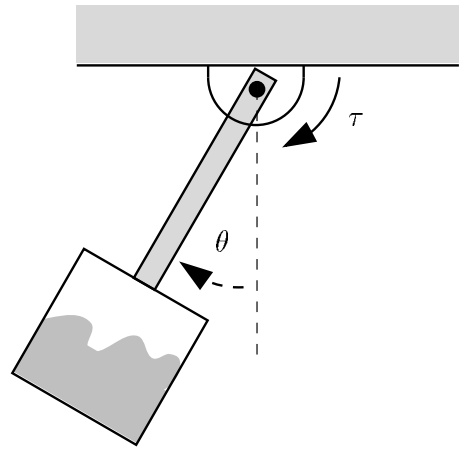}
\caption{An example input-output system. \label{fig:pendulum}}
\end{figure}

In many physics settings one is more concerned with the macroscopic
behavior of a large system, and less concerned with the system's
microscopic details.  As an admittedly contrived example, consider a
pendulum in a plane at whose free end is a tank partially filled with a
classical fluid (see Fig.~\ref{fig:pendulum}).  Suppose that at time
$t=0$ the system is at rest in its stable equilibrium, and our only
method of disturbing the system is to exert a time-varying torque
$\tau(t)$ at the pivot.  Suppose further that we are concerned only with
the time evolution of pendulum's angle $\theta(t)$ --- not with the
distribution of the fluid, given by some high-dimensional variable
$\Phi(t)$.

An exact model including the full fluid state $\Phi$ would give the
system's exact response to a driving torque, but would be quite
impractical.  As we only wish to describe the angular response to the
driving torque (a mapping from one degree of freedom to another) one
suspects a lower-dimensional model might suffice.  For example, one
might try treating the fluid as a point mass attached to the pendulum by
a non-linear spring.  

\begin{figure}
\includegraphics{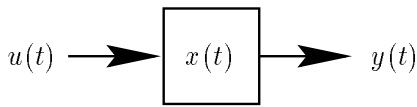}
\caption{A block diagram.\label{fig:blockdiagram}}
\end{figure}

Systems of this sort are naturally phrased in a control theory language.
In the typical control scenario, a time-varying input $u(t)$ drives a
system with state $x(t)$ giving output signal $y(t)$, and the system
dynamics have the form
\begin{equation}
\label{eqn:dynamics}
\begin{array}{rcl}
\dot{x}(t) &=& f(x(t),u(t)) \\
y(t) &=& h(x(t),u(t)).
\end{array}
\end{equation}
Such systems are depicted by a block diagram as shown in
Fig.~\ref{fig:blockdiagram}.  In the example of the pendulum, the
system's state is given by $x = (\theta, \dot{\theta},\Phi,\dot{\Phi})$
so as to describe the evolution with the first-order dynamics
(\ref{eqn:dynamics}).  The input to the system is $u(t) = \tau(t)$, and
the system's output is $y(t) = \theta(t)$.
 
Together with some initial condition (typically $x(0)=0$), the functions $f$
and $h$ in (\ref{eqn:dynamics}) define an \textit{input-output map}
$\Psi$ taking $u$ to $y$.  Since it is this relation with which we are
concerned, rather than the system's internal dynamics, a theoretical
model for the system will suffice if it describes $\Psi$.
Given some system of the form (\ref{eqn:dynamics}), model reduction aims to
produce simpler models (i.e. models with a lower-dimensional state $x$)
that approximate the original input-output map.

We will consider models of the form
\begin{equation}
\label{eqn:linear}
\begin{array}{rcl}
\dot{x} &=& Ax+Bu \\
y &=& Cx+Du
\end{array}
\end{equation}
where $x \in \mathbb{R}^n$, $u \in \mathbb{R}^m$, and $y \in
\mathbb{R}^p$, and $A$, $B$, $C$ and $D$ are time-independent real
matrices of sizes $n\times n$, $n \times m$, $p \times n$ and $p \times
m$ respectively.  (This entire discussion also holds for
complex-valued systems.)  Such linear models are called
\textit{state-space models}, and the \textit{order} of a model is $n$,
the dimension of the state $x$.  For compactness, the model with
matrices $A$, $B$, $C$ and $D$ is denoted by
\begin{equation}
\left (
\begin{array}{c|c}
A & B \\
\hline
C & D
\end{array}
\right ).
\end{equation}
(This notation should not be confused with the $(n+p)\times(n+m)$ matrix
with blocks $A$, $B$, $C$ and $D$.)  Often the output will only depend
on the system state, i.e.~$D=0$.  

Consider the dynamics (\ref{eqn:linear}) under a change-of-basis $z =
Tx$, where $T$ is invertible but need not be unitary.  The
dynamics may then be written as
\begin{equation}
\begin{array}{rcl}
\dot{z} &=& TAT^{-1}z+TBu \\
y &=& CT^{-1}z+Du.
\end{array}
\end{equation}
Thus changing the basis of the state space defines a mapping on 
state-space models given by
\begin{equation}
\left ( \begin{array}{c|c}
A & B \\
\hline
C & D
\end{array} \right )
\; \mapsto \; 
\left ( \begin{array}{c|c}
TAT^{-1} & TB \\
\hline
CT^{-1} & D
\end{array} \right ).
\end{equation}
(Note that $D$ is unchanged.)  We will call such maps \textit{similarity
transformations}.  Because these transformations are merely a rewriting
of the system dynamics, the input-output map remains the same.

\begin{figure}
\includegraphics{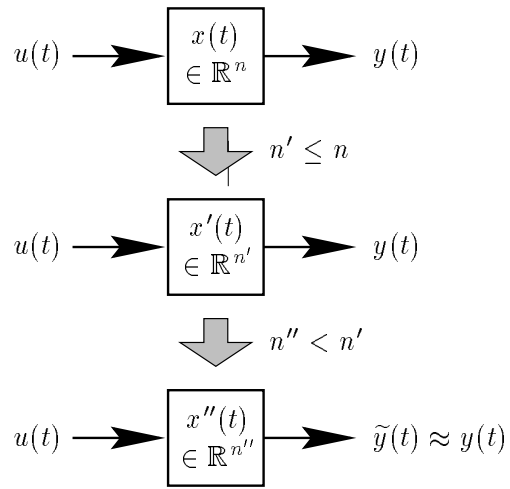}
\caption{Given an order $n$ model, we first reduce the system to the
lowest order $n' \leq n$ with the same input-output map, and then reduce
to an approximate model of order $n'' < n'$.\label{fig:procedure}}
\end{figure}

As a given state-space model is not a unique description of an
input-output map $\Psi$, in the next section we ask: what is the
lowest-order model with the same $\Psi$ as the given model?  After
finding a lowest-order exact model, we will use balanced truncation to
find lower-order models approximating $\Psi$.  This procedure is
summarized in Fig.~\ref{fig:procedure}.

\section{Lowest-Order Exact Models \label{sec:minreal}}
To find state-space models of lowest order with the exact $\Psi$ of a
given model, we will split the problem into two parts:
\textit{controllability} and \textit{observability}.  We will then
combine these ideas to find \textit{minimal realizations} of the system.

\subsection{Controllability}
Assume the system is initially in the state $x(0) = 0$; given a time
$\tau>0$, the \textit{controllable} states $x_f$ are those for which
there is an input signal $u(t)$ yielding $x(\tau) = x_f$.  The dynamics
(\ref{eqn:linear}) can be integrated to yield
\begin{equation}
\label{eqn:input}
x(\tau) = \int_0^\tau e^{A(\tau-t)}Bu(t)dt,
\end{equation}
which gives a linear map $\Psi^{(\tau)}_c:u\rightarrow x(\tau)$.  The
controllable states form the image of this map.  Since the map is
linear, the image is a subspace of the state-space $\mathbb{R}^n$,
called the \textit{controllable subspace}.  We denote this subspace by
$\mathcal{C}_{AB}$, as controllability only depends on the matrices
$A$ and $B$.

Using properties of the matrix exponential it can be shown that the
controllable subspace is the image of the
\textit{controllability matrix}:
\begin{equation}
\label{eqn:con_mat}
\mathcal{C}_{AB} = \mathrm{Image}\,
[B \; AB \; A^2 B \; \ldots \; A^{n-1}B].
\end{equation}
Thus we see that $C_{AB}$ is independent of $\tau$.  It can also be
shown that the controllable subspace is the image of the
\textit{controllability gramian}, an $n \times n$ matrix given by
\begin{equation}
\label{eqn:con_gram}
X_C = \int^\tau_0 e^{At}BB^\dagger e^{A^\dagger t} dt,
\end{equation}
and the orthogonal subspace of uncontrollable states is given by the
controllability gramian's kernel.  If $C_{AB} = \mathbb{R}^n$
(i.e. there exists an input signal $u$ to prepare any state $x(\tau)$)
then we say that $(A,B)$ is a \textit{controllable pair}.

It can be shown that given any $A$ and $B$, we can find a similarity
transformation such that the transformed matrices have the block
structure
\begin{equation}
\label{eqn:con_separation}
\wt{A} = TAT^{-1} = 
\left [ \begin{array}{cc}
\wt{A}_{11} & \wt{A}_{12} \\
0 & \wt{A}_{22}
\end{array} \right ]
\;\;\;\;\;
\wt{B} = TB = 
\left [ \begin{array}{c}
\wt{B}_{1} \\
0 
\end{array} \right ]
\end{equation}
with $(\wt{A}_{11},\wt{B}_1)$ a controllable pair.  Writing the state
vector as $x = (x_1,x_2)$ corresponding to this block structure,
we have
\begin{equation}
\begin{array}{rcccccl}
\dot{x}_1 & = & \wt{A}_{11}x_1& + & \wt{A}_{12}x_2& + & \wt{B}_1 u \\
\dot{x}_2 & = &               &   & \wt{A}_{22}x_2.
\end{array}
\end{equation}
Because $x(0) = 0$, these dynamics yield $x_2(t) = 0$ for all time.
Thus the dynamics for $x_1$ reduce to
\begin{equation}
\begin{array}{rcl}
\dot{x}_1 & = & \wt{A}_{11}x_1 + \wt{B}_1 u.
\end{array}
\end{equation}
Because $(\wt{A}_{11},\wt{B}_1)$ is a controllable pair, we may choose an
input $u$ to prepare any state $x_1(\tau)$, and thus the transformed
controllable subspace $\mathcal{C}_{\wt{A}\wt{B}}$ is given by the
states of the form $(x_1,0)$.  The orthogonal subspace, given by states
of the form $(0,x_2)$, is irrelevant to the input-output map since no
input can affect these states.

\subsection{Observability}

We now consider another problem with a similar structure.  Suppose the system
is in some initial state $x(0) = x_0$ and $u=0$ for all time.  Based on
the output $y(t)$ for $0 \leq t \leq \tau$, can we uniquely identify
$x_0$?  Integrating the dynamics (\ref{eqn:linear}) yields
\begin{equation}
\label{eqn:output}
y(t) = Ce^{At}x_0,
\end{equation}
which gives a linear map $\Psi^{(\tau)}_o: x_0 \rightarrow y$ (where by
$y$ we mean the output signal $y(t)$ for $0 \leq t \leq \tau$).
Suppose two initial states $x_0$ and $x_1$ give the same $y$.  As 
$\Psi^{(\tau)}_o$ is linear, the initial state $x_0 - x_1$ must give $y=0$.
We call initial states giving output $y=0$ \textit{unobservable}, since
any unobservable state may be added to any other initial state without
changing the output.

The unobservable states form the kernel of $\Psi^{(\tau)}_o$;
as the map is linear these states form a subspace, called the
\textit{unobservable subspace} and denoted by $\mathcal{N}_{CA}$.
It can be shown that the unobservable subspace is given by the kernel of
the \textit{observability matrix}:
\begin{equation}
\label{eqn:obs_mat}
\mathcal{N}_{CA} = 
\mathrm{ker}\,\left [ \begin{array}{c}
C \\
CA \\
\vdots \\
CA^{n-1}
\end{array}
\right ].
\end{equation}
Thus $\mathcal{N}_{CA}$ is independent of $\tau$.
It can also be shown that $\mathcal{N}_{CA}$ is given by the kernel of the
\textit{observability gramian}
\begin{equation}
\label{eqn:obs_gram}
Y_O = \int_0^\tau e^{A^\dagger t}C^\dagger C e^{At} dt,
\end{equation}
and the orthogonal subspace of observable states is given by the
observability gramian's image.  If $\mathcal{N}_{CA} = \{\vec{0}\}$,
then the entire space is observable, and we say that $(C,A)$ is an
\textit{observable pair}.  As the observable states are given by the
image of (\ref{eqn:obs_gram}) and the controllable states are given by
the image of (\ref{eqn:con_gram}), $(C,A)$ is an observable pair if and
only if $(A^\dagger,C^\dagger)$ is a controllable pair.

It can be shown that given any $C$ and $A$, we can find a similarity
transformation such that the transformed matrices have the block
structure
\begin{equation}
\label{eqn:obs_separation}
\begin{array}{rcccl}
\wt{A} &=& TAT^{-1} &=& 
\left [ \begin{array}{cc}
\wt{A}_{11} & 0 \\
\wt{A}_{21} & \wt{A}_{22}
\end{array} \right ]
\\
\\
\wt{C} &=& CT^{-1} &=& 
\left [ \begin{array}{cc}
\,\;\wt{C}_{1}\,\; & \;0\,\; 
\end{array} \right ]
\end{array}
\end{equation}
with $(\wt{C}_1,\wt{A}_{11})$ an observable pair.  Writing the state
vector as $x = (x_1,x_2)$ corresponding to this block structure,
we have
\begin{equation}
\begin{array}{rccccl}
\dot{x}_1 & = & \wt{A}_{11}x_1 \\
\dot{x}_2 & = & \wt{A}_{21}x_1 & +  & \wt{A}_{22}x_2 \\
        y & = & \wt{C}_1 x_1. 
\end{array}
\end{equation}
Thus the time evolution of $x_1$ is never affected by $x_2$, and the
output signal $y$ only depends on $x_1$.  Because
$(\wt{C}_{1},\wt{A}_{11})$ is an observable pair, we can uniquely
identify an initial state $x(0) = (x_1,0)$ based on $y$.  The
transformed unobservable subspace $\mathcal{N}_{\wt{C}\wt{A}}$ is given
by the states of the form $(0,x_2)$, and is irrelevant to the
input-output map since no output can be affected by these states.

\subsection{Minimal Realizations}
The notions of controllability and observability give us a means of
deciding whether a state affects the system's input-output map: if a
state is unobservable, it does not affect the output, and if a state is
uncontrollable, it is unaffected by the input.  Only those states that
are both controllable and observable are of relevance.

We say that a state-space model given by matrices $(A,B,C,D)$ is a
\textit{minimal realization} if no lower-order model gives the same
input-output map $\Psi$.  The intuition above can be made precise as
follows: it can be shown that a model is a minimal realization if and
only if all states are both controllable and observable, i.e. $(A,B)$ is
a controllable pair and $(C,A)$ is an observable pair.

Given a state-space model given by $A$, $B$, $C$ and $D$ we may find a
minimal realization by isolating only those dimensions which are both
controllable and observable.  To do so, we perform a \textit{Kalman
decomposition}, which simultaneously performs the transformations
(\ref{eqn:con_separation}) and (\ref{eqn:obs_separation}) as follows.
For any state-space model there exists a similarity transformation such
that the matrices of transformed model have the block structure
\begin{equation}
\left ( \begin{array}{c|c}
TAT^{-1} & TB \\
\hline
CT^{-1} & D
\end{array} \right )
\; = \;
\left (\begin{array}{cccc|c}
\wt{A}_{11} &    0        & \wt{A}_{13} & 0           & \wt{B}_1\\
\wt{A}_{21} & \wt{A}_{22} & \wt{A}_{23} & \wt{A}_{24} & \wt{B}_2 \\
0           & 0           & \wt{A}_{33} & 0           & 0 \\
0           & 0           & \wt{A}_{43} & \wt{A}_{44} & 0 \\
\hline
\wt{C}_1    & 0           & \wt{C}_3    & 0           & D     
\end{array} \right )
\end{equation}
and, writing the state vector as $x = (x_1,x_2,x_3,x_4)$ corresponding
to the block structure, the controllable states are of the form
$(x_1,x_2,0,0)$ and the observable states are of the form
$(x_1,0,x_3,0)$.   Thus only states of the form $(x_1,0,0,0)$ are both
controllable and observable.

Eliminating all but the states $(x_1,0,0,0)$ yields the state-space model
\begin{equation}
\left ( \begin{array}{c|c}
\wt{A}_{11} & \wt{B}_1 \\
\hline
\wt{C}_1 & D
\end{array} \right ).
\end{equation}
Since we have only eliminated states irrelevant to the input-output map,
this reduced system has the exact same $\Psi$ as the original system.
Further, since states of the form $(x_1,0,0,0)$ were both controllable
and observable, $(\wt{A}_{11}, \wt{B}_1)$ is a controllable pair and
$(\wt{C}_1, \wt{A}_{11})$ is an observable pair; thus this model is a
minimal realization.  

Sometimes the uncontrollable and unobservable states are obvious from
the form of a physical system's dynamics, e.g.~some degrees of freedom
are uncoupled, or some symmetry can be exploited.  In other
circumstances, the physics may not make it clear which states can be
eliminated without affecting the input-output map.  Especially in the
latter situation, an algorithmic method such as the Kalman decomposition
(available in MATLAB) can be quite advantageous.  Nonetheless, one feels
intuitively that one should always be able to find a minimal
realization analytically if one is sufficiently clever.  However, we do
not expect in general that standard analytic methods will be useful in
seeking lower-order approximations, as we will do in the next section.

\section{Lower-Order Approximate Models\label{sec:baltrunc}}
To apply balanced truncation, we will first assume that we have reduced
the model to a minimal realization.  We make the further assumption
that the resulting system is exponentially stable, i.e.~all eigenvalues
of $A$ have strictly negative real part.  (Various methods exist
to extend these methods to unstable systems, e.g., \cite{Sznaier},
but we assume stability to prove the standard result.)

\subsection{Quantifying Observability and Controllability}

In the previous section we distinguished between states that were
observable and unobservable; we now wish to quantify the observability
of the observable states.  To do so, consider the output signal
for $t \geq 0$ that results from the initial state $x(0)=x_0$ when there
is no input $u$.  This signal is given by (\ref{eqn:output}).  Using the
$\mathcal{L}_2$ norm for the signal $y$, we have
\begin{equation}
\label{eqn:obs_norm}
\begin{array}{rcl}
||y||^2 &=& \int^\infty_0 y^\dagger(t)y(t)dt \\
        &=& x_0^\dagger \left (
             \int^\infty_0 e^{A^\dagger t}C^\dagger C e^{At} dt
              \right ) x_0 \\
        &=& x_0 Y_O x_0
\end{array}
\end{equation}
where $Y_O$ is the observability gramian given by (\ref{eqn:obs_gram})
with $\tau \rightarrow \infty$.  Recall that the kernel of $Y_O$ is
independent of $\tau$, so choosing $\tau \rightarrow \infty$ will not
change which states are observable and which states are
unobservable. (System stability is required to ensure convergence of the
integral in this limit.)  Scaling the norm-squared of the output by the
norm-squared of the initial state yields
\begin{equation}
\frac{||y||^2}{||x_0||^2} = \frac{x_0^\dagger Y_O x_0}{x_0^\dagger
x_0},
\end{equation}
which quantifies the observability of the states in the
direction of $x_0$.  

\begin{figure}
\includegraphics[scale=0.75]{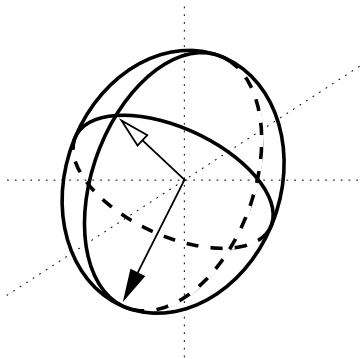}
\caption{An observability ellipsoid.  The solid arrow points in a
direction more observable than the direction of the open
arrow.\label{fig:ellipsoid}}
\end{figure}

From (\ref{eqn:obs_norm}) we see that the observability gramian is
Hermitian and positive semidefinite.  As we have a minimal realization,
all states $x_0$ are observable ($x_0 Y_O x_0 > 0$), and so $Y_O$ is
strictly positive definite.  Thus in geometric terms analogous to the
moment-of-inertia tensor, $Y_O$ defines an ``observability ellipsoid''
in the state space, with the longest principal axes along the most
observable directions (see Fig.~\ref{fig:ellipsoid}). A similarity
transformation given by $T$ transforms the observability gramian by
$Y_O \rightarrow (T^{-1})^\dagger Y_O T^{-1}$.  As $T$ need not be
unitary, this transformation may rescale the ellipsoid's axes as well
as rotate them.

We quantify controllability in a similar fashion.  Suppose that at a
time well in the past ($\tau \rightarrow -\infty$) the system is in the
state $x=0$, and some input $u(t)$ drives the system for $t \leq 0$,
yielding a final state $x(0) = x_0$.  As we have a minimal realization,
all states $x_0$ are controllable and therefore can be prepared in this
fashion.  For each state $x_0$ there is a minimum signal size
$||u_\mathrm{opt}||^2 = \int_{-\infty}^0 u(t)^\dagger u(t)dt$ required to
yield $x(0)=x_0$.  The smaller this minimum signal, the more
sensitive this state is to the input signal.  Thus states with a smaller 
$||u_\mathrm{opt}||^2$ are said to be more controllable.

Consider the controllability gramian $X_C$ given by (\ref{eqn:con_gram})
with $\tau \rightarrow \infty$. Just as $Y_O$, $X_C$ is Hermitian and
positive semidefinite, and because we have assumed a minimal
realization, $X_C$ is strictly positive definite and therefore
invertible.  It can be shown that
\begin{equation}
||u_\mathrm{opt}||^2 = x_0^\dagger X_C^{-1}x_0.
\end{equation}
It follows that
\begin{equation}
\left ( \frac{||u_\mathrm{opt}||^2}{||x_0||^2} \right )^{-1} 
=
\frac{x_0^\dagger X_C x_0}{x_0^\dagger x_0},
\end{equation}
which quantifies the controllability of the states in the direction of
$x_0$.  Just as with $Y_O$, $X_C$ defines a ``controllability
ellipsoid'' in state space, with the longest principal axes along the
most controllable directions.  A similarity transformation given by $T$
transforms the controllability gramian by $X_C \rightarrow T
X_C T^\dagger$.

We have thus found that the gramians give us a useful measure of a
state's observability and controllability.  One might ask whether the
observability and controllability matrices of (\ref{eqn:obs_mat}) and
(\ref{eqn:con_mat}) could serve a similar purpose, but in fact they are
only useful for determining \textit{whether} a given state is
observable/controllable.  One might also ask if $\tau \rightarrow
\infty$ is necessary --- we could choose finite $\tau$, and quantify
controllability and observability on a finite time horizon.  However,
finite $\tau$ does not lead to the error bound in the approximations
of the next section, which is the main result of
these notes.

\subsection{Balanced Truncation}
With the above quantification of observability and controllability, one
might be tempted to prescribe some algorithm like eliminating the least
observable or least controllable dimensions in the state space to yield
a lower-order approximate model.  However, such an approach would not
necessarily be successful.  Suppose, for example, that the least
observable states were in the direction of the unit vector $\hat{x}$,
but that states in this direction were extremely controllable.  Thus a
small signal $u$ might lead to the internal state $x=\lambda \hat{x}$ with
$\lambda$ large. Though this state is the least observable, $\lambda$
might be sufficiently large that the resulting output signal is
non-negligible.

\begin{figure}
\includegraphics{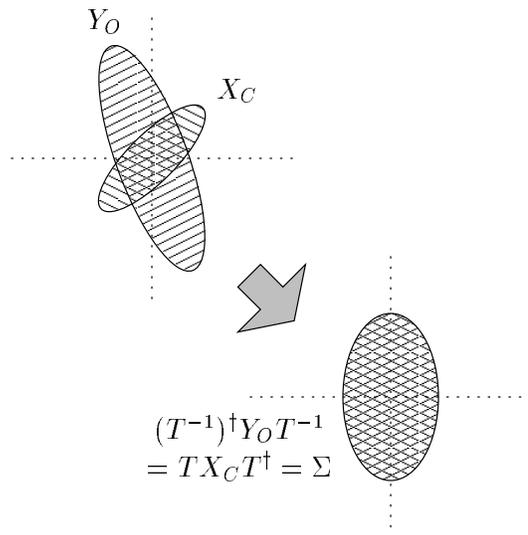}
\caption{The balancing transformation $T$ transforms the observability
and controllability ellipsoids to an identical ellipsoid aligned with
principle axes along the coordinate axes.\label{fig:balancing}}
\end{figure}

Instead, we wish to use observability and controllability to yield a
single measure of a state's importance to the input-output map $\Psi$.
It can be shown that given two positive definite square matrices $X_C$
and $Y_O$ of the same size, there exists an invertible $T$ such that
\begin{equation}
T X_C T^\dagger = (T^{-1})^\dagger Y_O T^{-1} = \Sigma
\end{equation}
where $\Sigma$ is diagonal with positive real diagonal entries.  Such a
similarity transformation is called a \textit{balancing
transformation}.  Geometrically, balancing transforms the observability
and controllability ellipsoids so that they are identical and their
principal axes lie on the coordinate axes of the state space (see
Fig.~\ref{fig:balancing}).

The resulting $\Sigma$ is unique up to permutation of the diagonal
elements, so we may choose $T$ yielding the transformed gramians
\begin{equation}
\widetilde{X}_C = \widetilde{Y}_O = \Sigma = \left [
\begin{array}{cccc}
h_1 \\
& h_2 \\
& & \ddots \\
& & & h_n 
\end{array}
\right ]
\end{equation}
with $h_1 \geq h_2 \geq \ldots \geq h_n > 0$.  The $h_i$ are called
\textit{Hankel Singular Values} (HSVs); in this transformed system $h_i$
is the quantitative measure of both the observability and
controllability of the unit basis vector $\hat{e}_i$.  Thus the basis
vectors have been sorted in order of relevance to the input-output map.

Once the system is balanced, we may truncate the state-space dimensions
with low HSVs to yield lower-order approximate models. Intuitively, the
smaller the HSVs corresponding to truncated dimensions, the better the
approximation.  We will now make this idea precise.

Let the balanced state-space model with sorted HSVs be given by matrices
$\wt{A}$, $\wt{B}$, $\wt{C}$ and $D$.  Choosing some $r$ such that $h_r$
is strictly greater than $h_{r+1}$, we write the state vector $x =
(x_1,x_2)$ where $x_1$ gives the first $r$ coordinates, and $x_2$ gives
the last $n-r$ coordinates.  We then write the state-space model
in the corresponding block structure
\begin{equation}
\label{eqn:original}
\left ( \begin{array}{c|c}
\wt{A} & \wt{B} \\
\hline
\wt{C} & D
\end{array} \right )
\; = \;
\left (\begin{array}{cc|c}
\wt{A}_{11} & \wt{A}_{21} & \wt{B}_1 \\
\wt{A}_{12} & \wt{A}_{22} & \wt{B}_2 \\
\hline
\wt{C}_1    & \wt{C}_2    & D
\end{array} \right )
\end{equation}
and truncate the $n-r$ least significant dimensions of the model, yielding
the order $r$ model
\begin{equation}
\label{eqn:truncated}
\left (
\begin{array}{c|c}
\wt{A}_{11} & \wt{B}_1  \\
\hline
\wt{C}_2  & D 
\end{array}
\right ).
\end{equation}
This procedure is called \textit{balanced truncation}.

We now give bounds on the resulting approximation error.  Let $u(t)$ be
some input signal on $t \in (-\infty,\infty)$ that is finite with
respect to the $\mathcal{L}_2$ norm
\begin{equation}
\label{eqn:L_2_norm}
||z||^2 = \int^\infty_{-\infty} \!\!\!z(t)^\dagger z(t) dt.
\end{equation}
Let $y(t)$ and $\wt{y}(t)$ be the resulting output signals of the
original and truncated systems respectively.  Stability of the original
system (\ref{eqn:original}) and the strict inequality $h_r > h_{r+1}$
guarantees stability of the truncated system (\ref{eqn:truncated});
since both systems are stable, the output signals $y$ and $\wt{y}$ are
also finite with respect to the norm (\ref{eqn:L_2_norm}), as is the
error signal $\wt{y} - y$.  Now, letting 
$h^\mathrm{tr}_1 > h^\mathrm{tr}_2 > \ldots > h^\mathrm{tr}_k$
be the
\textit{distinct} HSVs of the truncated $n-r$ dimensions, it can be
shown that
\begin{equation}
\label{eqn:bound}
h^\mathrm{tr}_1 \leq \max_u \frac{||\wt{y}-y||}{||u||} \leq 
2 \sum_{j=1}^k h^\mathrm{tr}_j
\end{equation}
Thus we have both an upper and lower bound on the worst-case errors
resulting from balanced truncation.  Given the Hankel Singular Values of
a system, we may truncate to a system of desired order and bound the
resulting error, or we may choose the order of truncation based on the
maximum tolerable error.

The advantages of balanced truncation as method of approximating the
input-output map are two-fold.  First, the error is bounded as above, in
contrast to more traditional approximations where errors estimated by
the order of some small parameter, e.g. $O(\epsilon^2)$.  Second, the
algorithmic process is efficient.  Balanced truncation does not
necessarily yield the \textit{optimal} order-$r$ approximation in the
sense of the error bound (\ref{eqn:bound}), but finding the optimal
approximation may be computationally difficult.

For single-input single-output systems (i.e. $u$, $y \in \mathbb{R}$), the
error bound may also be described in terms of frequency responses.
Given a sinusoidal input, a state-space model will give an output
with the same frequency and a frequency-dependent amplitude and phase
shift:
\begin{equation}
u(t) = \sin(\omega t) \; \rightarrow \; y(t) = A_\omega \sin(\omega t +
\phi_\omega).
\end{equation}
Let $A_\omega$ and $\phi_\omega$ be the amplitude and phase shifts for
the exact system, and let $\wt{A}_\omega$ and $\wt{\phi}_\omega$ be the shifts
for the approximate system.  Writing the exact and approximate shifts
in complex form, we have
\begin{equation}
h^\mathrm{tr}_1 \leq 
\max_\omega \left |\wt{A}_\omega e^{i\wt{\phi}_\omega} - A_\omega
e^{i\phi_\omega} \right |
\leq 
2 \sum_{j=1}^k h^\mathrm{tr}_j.
\end{equation}
Thus even though sinusoidal inputs and outputs are unbounded in the
$\mathcal{L}_2$ norm (\ref{eqn:L_2_norm}), the error is controlled 
in this fashion.

\section{Conclusion}

According to \cite{dullerud}, balanced realizations first appeared in
the control literature in 1981, and the proof of the error bound on
truncated models first appeared in 1984.  Until recently, however, the
physics community has made little use of this powerful tool.  We believe
that balanced truncation will be of particular value when building
simulations of and theoretical models for the evolution of
macroscopic quantities in large complex systems, and it is hoped
that these notes will be helpful in such studies.

\end{document}